\documentclass[a4paper]{jpconf}
\pdfoutput=1
\usepackage{graphicx}
\usepackage{amsmath}
\usepackage{amssymb}
\usepackage{enumerate}
\usepackage{color}

\begin{document}
\title{Topography effect on the seismogenic deformation of the earth's
surface}

\author{Gustavo Lara$^{1 *} $, Gabriel \'Alvarez$^1$, Gabriel Gonz\'alez$^2$,
     Juan Gonz\'alez$^2$, Rafael Ar\'anguiz$^3$ and Patricio Catal\'an$^4$}

\address{$^1$ Universidad de Antofagasta, Av. Angamos 601, Antofagasta, Chile.
\\ $^2$ Universidad Cat\'olica del Norte, Av. Angamos 610, Antofagasta, Chile.
\\ $^3$ Universidad Cat\'olica de la Ssma. Concepci\'on, Alonso de
Ribera 2850, Concepci\'on,  Chile.
\\ $^4$ Universidad T\'ecnica Federico Santa Mar\'{\i}a, Av.
Espa\~{n}a 1680,  Valpara\'{\i}so, Chile. }

\ead{$^*$gustavo.lara@uantof.cl}

\begin{abstract}
A comparison of the displacements of the earth's surface after an earthquake was
made, calculating with the analytical expressions coming from an infinite flat
slab approximation and compared with these numerically considering
the topography of the Earth.
One conclusion of this work is that the flat Earth approximation, has a greater
error in the lateral displacement
than in the vertical one. It can also be noted that the error in the magnitude
of the displacement is less or of the order of ten percent of the maximum
displacement of the earth's surface.
\end{abstract}

\section{Introduction}
In some fields of Geophysics it is necessary to know the deformation of the
surface of the Earth due to the slip in geological faults. It's known the
analytical solutions \cite{Mansinha1971,Okada1985}, but where the Earth
is considered an infinite flat slab. Currently
the effect of introducing real topography approximations has been considered in
the literature. General solutions to this problem are numerical. There are
examples of these estudies \cite{Armigliato2003,Lin2013a,Lin2013b}.
In this contribution, we develop a numerical
method based on the boundary element method, which is applied to a
homogenous, isotropic Earth and with the real topography of its surface.

\section{Model for the Earth}
The Earth is considered to be a linear, homogeneous, isotropic elastic solid of
ellipsoidal geometry described by WGS 84. The geological fault is represented by
a cutting surface, $\Sigma$, where the abrupt dislocation of one face with
respect to the other, tangent to the fault plane, gives rise to the earthquake
and the deformation of the external terrestrial surface, $S$.

From the equations that describe this type of solid we can deduce the
displacement, ${\bf u}$, of any point ${\bf x}$ on the external surface S,
only knowing the displacements in the points ${\bf y}$ of the entire surface

\begin{equation}
\begin{aligned}
{\bf u} ({\bf x}) = &
\int_{\Sigma}
\frac{a}{r^2}
\Big\{ -3{\bf \hat{r}}
\left( {\bf \hat{r}} \cdot {\bf U}\right)
 \left( {\bf \hat{r}} \cdot d{\bf \Sigma } \right)
-b \big[
{\bf U} \left( {\bf \hat{r}} \cdot d{\bf \Sigma}\right)
+  \left( {\bf \hat{r}} \cdot {\bf U}\right) d{\bf \Sigma}
\big]
\Big\}
\\ &
\!\!- \!  {\cal P} \int_{S}
\frac{a}{r^2}
\Big\{ {\bf \hat{r}}
\big[
-3\left( {\bf \hat{r}} \cdot {\bf u}\right)
 \left( {\bf \hat{r}} \cdot d{\bf S } \right)
+b \, {\bf u} \cdot d{\bf S}
\big]
-b \big[
{\bf u} \left( {\bf \hat{r}} \cdot d{\bf S}\right)
+  \left( {\bf \hat{r}} \cdot {\bf u}\right) d{\bf S}
\big]
\Big\}
\end{aligned}
\end{equation}
where $ {\bf r} = {\bf y} - {\bf x} = r \hat{\bf r} $,
$a=1/\left[ 4\pi (1-\nu) \right] $, $b=1-2\nu$
and $\nu $ is the Poisson's coefficient.

This integral equation is solved self-consistently using an irregular triangular
network (TIN) to represent the elements of the earth's surface. For the
creation of this TIN, we used the real topography of the Earth provided by Gebco
\cite{Gebco}.

The direction and magnitude of the dislocations in the fault, ${\bf U}$,  are
considered as input parameters and are obtained from slip models for the fault.
There is usually more than one slip model for each earthquake, however, here
we use a model for each of the cases  with the propose of revealing the
effect of the topography and without questioning the accuracy of the slip
model.

For each of the earthquakes above considered we choose a model of slip among
those published in the literature. We do not take into account the level of
accuracy of those models, since at present, we just want to compare the effect
of the real topography of the Earth in comparison of the flatness of the Earth
assumption.

\section{Results}

The infinite flat slab model and the real topography of the Earth are
compared, for four earthquakes of great magnitude.
The information that characterizes these earthquakes is shown in Table 1, where
 the maximum values of the difference between the two methods are also shown,
compared to the absolute value. These comparative values are made for the full
displacement vector ${\bf u}$, to its vertical component, $u_z$, and
to its lateral component $u_{_L}$.

\begin{center}
\begin{table}[h]
\caption{ Information about earthquakes.}
\centering
\begin{tabular}{@{}*{7}{lcccc}}
\br
Year & 2008 & 2010 & 2011 & 2014 \\
\mr
Location & Wenchuan & Maule & Tohoku-Oki & Iquique \\
      & China & Chile & Japan & Chile \\
\mr
Moment magnitude & 7.9 Mw & 8.8 Mw &  9.1 Mw & 8.1 Mw \\
\mr
Slip model & Fielding & Delouis & Hayes & Wei \\
       & 2013 \cite{Fielding2013} & 2010 \cite{Delouis2010}
       & 2011 \cite{Hayes2011} & 2014 \cite{Wei2014} \\
\mr
$ {\| \Delta {\bf u} \|_\text{max} } / {\| {\bf u} \|_\text{max}  } $
& ${0.58}/{5.07}$& ${0.72}/{6.69}$ & ${1.68}/{21.6}$ & ${0.13}/{1.57}$ \\
\mr
$ | \Delta u_z |_\text{max} / |u_z |_\text{max} $
& $ {0.22}/{4.73} $&${0.17}/{5.34}$&${1.15}/{9.73}$&${0.097}/{0.87}$ \\
\mr
$  | \Delta u_{̣_L} |_\text{max}  / {|u_{_L} |_\text{max}  } $
& ${0.58}/{3.05}$&$ {0.71}/{4.90}$ &$ {1.67}/{21.1}$&${0.12}/{1.49}$ \\
\br
\end{tabular}
\end{table}
\end{center}

From the values in Table 1 it can be seen that the Wenchuan and Maule
earthquakes have greater displacement magnitudes in vertical direction than in
lateral direction, however, the relative difference between the methods is
greater in the lateral direction than in the vertical direction.

For the earthquakes of Tohoku and Iquique it is just the opposite, they have
magnitudes of displacement greater in lateral direction more that in vertical
direction, and the relative difference between the methods, is
greater in vertical direction that in lateral direction.

Figures 1 to 4 correspond to a color maps for each of these earthquakes, which
represents the magnitude of the difference between the displacement considering
the real topography of the Earth and the displacement considering an infinite
flat slab. Additionally, height level curves (blue lines) are included that
allow visualization of the areas with height gradients (curves closer
to each other).
The green lines represent the level curves of the absolute displacement of the
earth's surface allowing visualizing that the biggest differences are not on the
regions with the greatest displacements.

\begin{figure}[h]
\begin{minipage}{17pc}
\includegraphics[width=17pc]{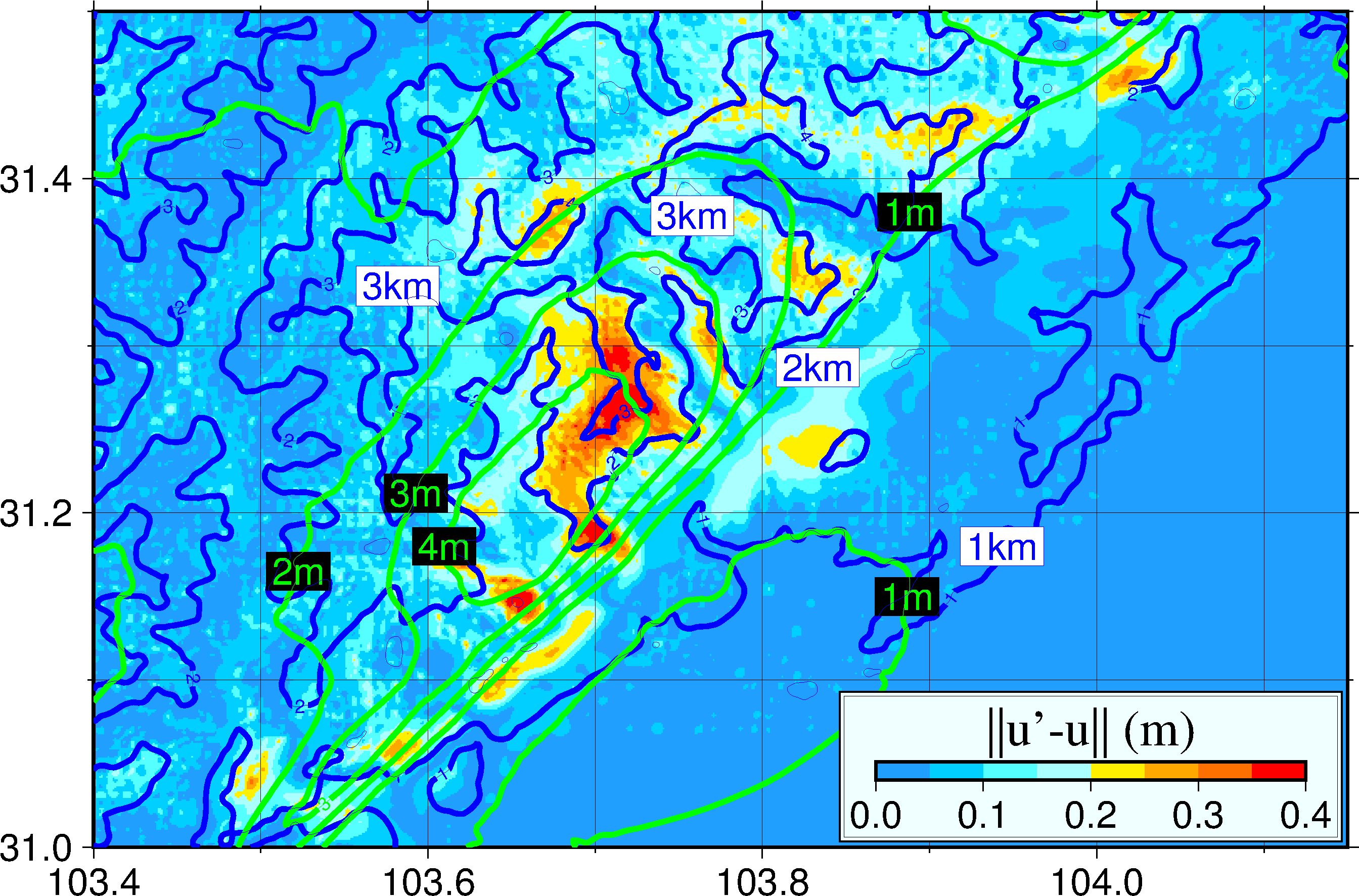}
\caption{\label{Figure1}Differences in displacements for earthquake 2008 Mw
7.9 Wenchuan (China).
The blue lines correspond to height level curves. The green lines correspond to
the displacement level curves.
}
\end{minipage}\hspace{2pc}%
\begin{minipage}{17pc}
\includegraphics[width=17pc]{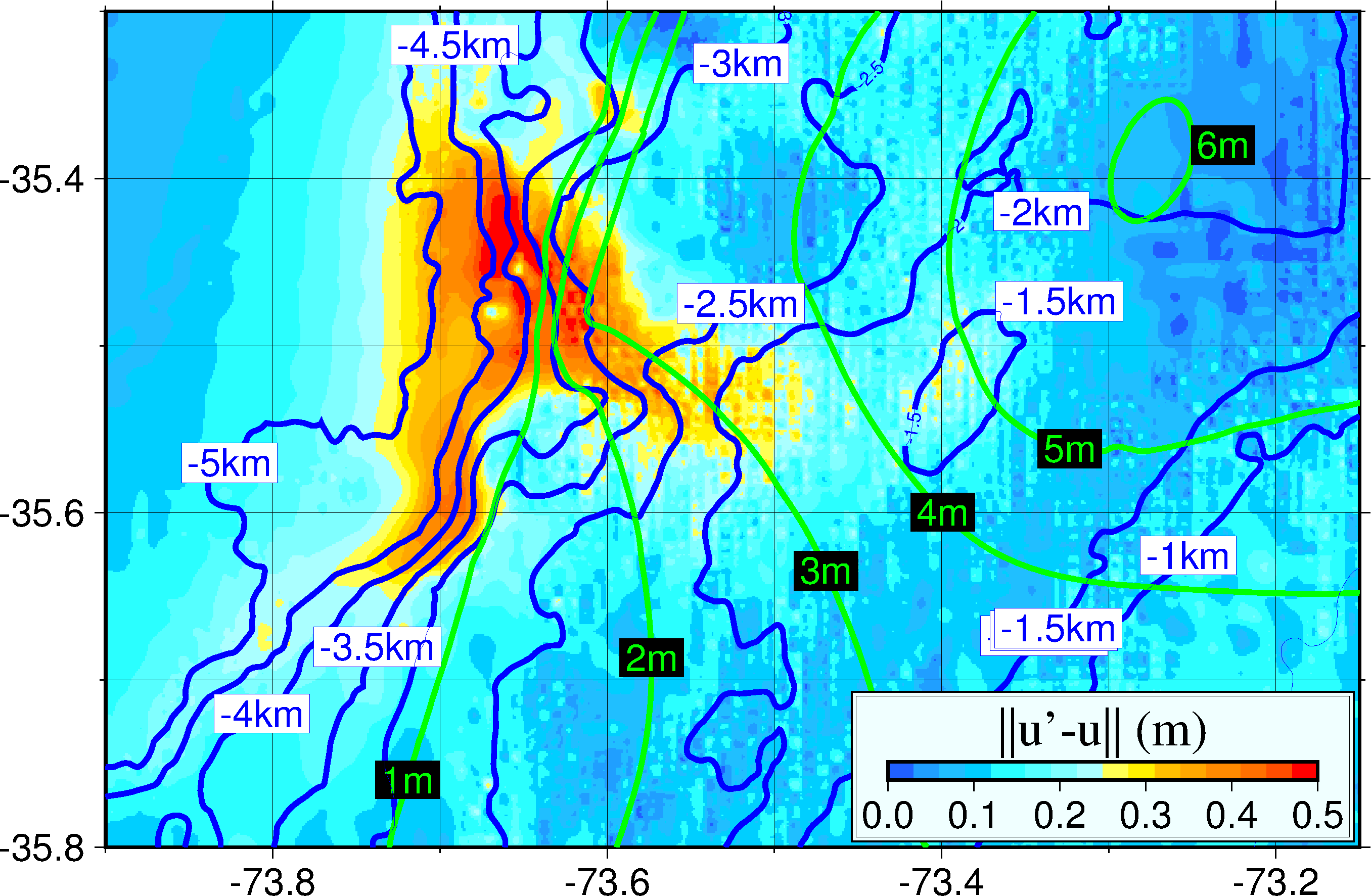}
\caption{\label{Figure2}Differences in displacements for earthquake 2010 Mw
8.8 Maule (Chile).
The blue lines correspond to height level curves. The green lines correspond to
the displacement level curves.
}
\end{minipage}
\end{figure}

\begin{figure}[h]
\begin{minipage}{17pc}
\includegraphics[width=17pc]{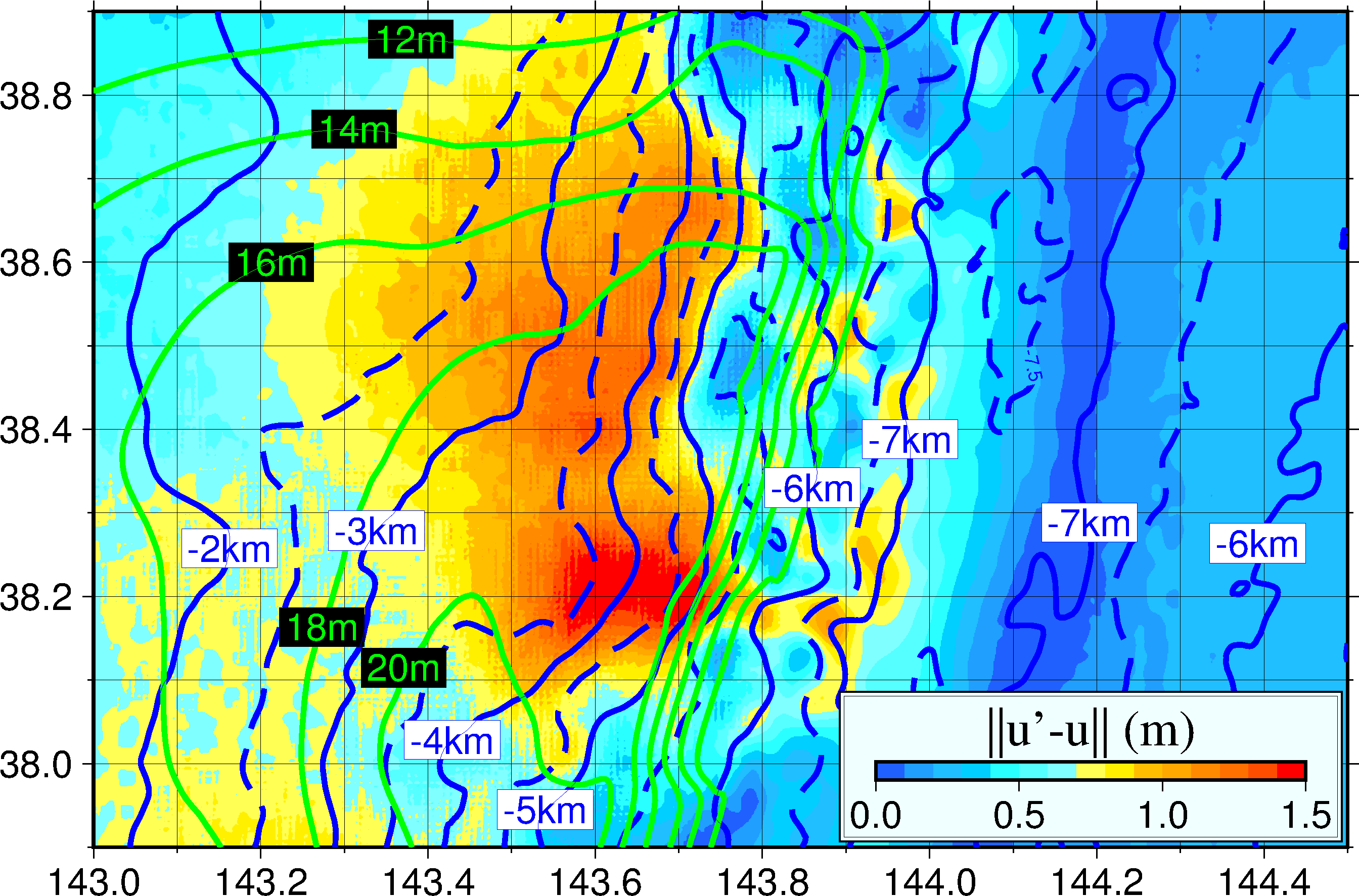}
\caption{\label{Figure3}Differences in displacements for earthquake 2011 Mw
9.1 Tohoku-Oki (Japan).
The blue lines correspond to height level curves. The green lines correspond to
the displacement level curves.}
\end{minipage}\hspace{2pc}%
\begin{minipage}{17pc}
\includegraphics[width=17pc]{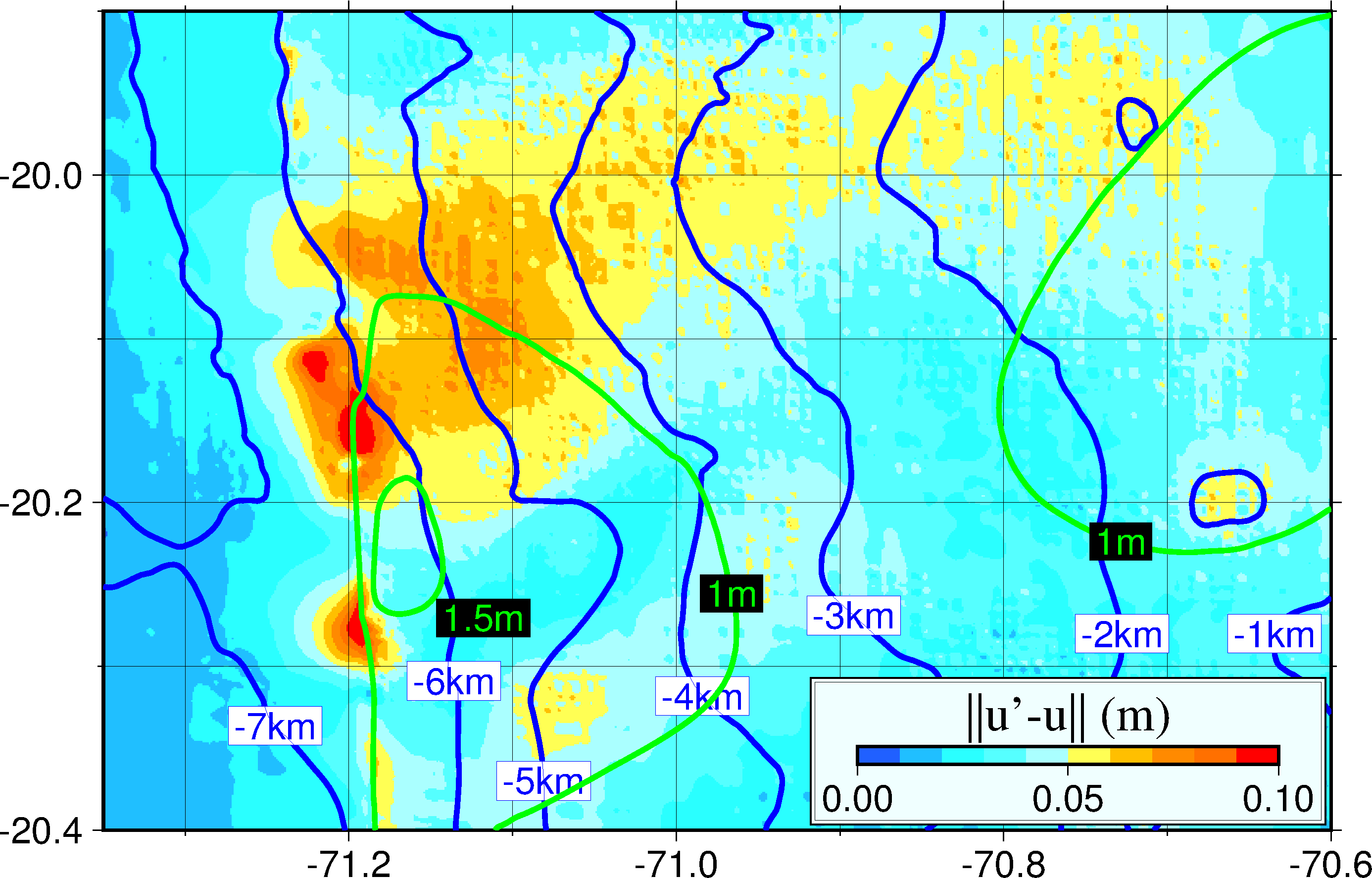}
\caption{\label{Figure4}Differences in displacements for earthquake 2014 Mw
8.1 Iquique (Chile). The blue lines correspond to height level curves. The green
lines correspond to
the displacement level curves.}
\end{minipage}
\end{figure}

\section{Conclusions}

\begin{enumerate}[(i)]

 \item This boundary element method allows to calculate of the
deformation on the external surface regardless of how irregular it is, using
known dislocations in a plane of failure. It's also allows the calculations
regardless of how irregular it is the plane of failure.

 \item The difference in the displacements is mostly lateral rather than
vertical.

 \item The magnitude of the difference in the displacements of these methods is
less or of the order of $10 \%$ of the maximum displacement over the entire
surface.

 \item The regions where there are greater differences between the methods are
also regions that show high gradients of height in their topography.

\end{enumerate}

\ack{ The authors thank financial support from Fondap project 15110017. }

\smallskip

\end{document}